\documentclass[a4paper]{jpconf}
\usepackage{graphicx}
\usepackage{amssymb,amsmath,amsthm}

\newcommand{\beqn}{\begin{equation}}
\newcommand{\eeqn}{\end{equation}}

\begin{document}
\title{Voices of Silence, Novelties of Noise:
Oblivion and Hesitation as Origins of Quantum Mysteries
}

\author{E Cohen$^1$ and A C Elitzur$^2$}

\address{$^1$School of Physics and Astronomy, Tel Aviv University, Ramat-Aviv, Tel Aviv 6997801, Israel}
\address{$^2$Iyar, The Israeli Institute for Advanced Research, Rehovot, Israel}

\ead{eliahuco@post.tau.ac.il}

\begin{abstract}
Among the (in)famous differences between classical and quantum mechanics, quantum counterfactuals seem to be the most intriguing. At the same time, they seem to underlie many quantum oddities. In this article, we propose a simple explanation for counterfactuals, on two levels. Quantum Oblivion (QO) is a fundamental type of quantum interaction that we prove to be the origin of quantum counterfactuals. It also turns out to underlie several well-known quantum effects. This phenomenon is discussed in the first part of the article, yielding some novel predictions. In the second part, a hypothesis is offered regarding the unique spacetime evolution underlying QO, termed Quantum Hesitation (QH). The hypothesis invokes advanced actions and interfering weak values, as derived first by the Two-State-Vector Formalism (TSVF). Here too, weak values are argued to underlie the familiar ``strong'' quantum values. With these, an event that appears to have never occurred can exert causal effects and then succumb to QO by another time-evolution involving negative weak values that eliminate them. We conclude with briefly discussing the implications of these ideas on the nature of time.
\end{abstract}

\section{Introduction} \label{sec:Introduction}

Seeking a suitable thank to the organizers of this unique conference and volume, we chose to present a nascent work in which rigor and speculation are hopefully well balanced. Accordingly, it has two parts:\\

i)	Quantum Oblivion (QO) refers to a very simple yet hitherto unnoticed type of quantum interaction, where momentum appears not to be conserved. The problem's resolution lies within a brief critical interval, during which more than one interaction takes place. Rapid self-cancellation, however, leaves only one interaction completed. While the paradox's resolution is novel, the interaction itself turns out to underlie several well-known quantum peculiarities. This gives a new, realistic twist to one of QM's most uncanny hallmark, namely, the {\it counterfactual}. Although rigorously derived from standard QM, Oblivion offers some new experimental predictions, as well as new insights into other quantum oddities. \\

ii)	Quantum Hesitation (QH) proceeds further to theorizing, as to {\it how} QO takes place. It reformulates Oblivion within time-symmetric interpretations of QM, mainly Aharonov's Two-State-Vector Formalism (TSVF). We assume that, beneath several momentary interactions, out of which only one is completed, there were several possible {\it histories}, which have left only one finalized. From the TSVF we then take one of its most exotic features, namely, weak values unbounded by the eigenvalue spectrum. This allows too-large/too-small/negative weak values appearing under special pre- and post-selections. Such ``unphysical'' values are assumed to evolve along both time directions, over the same spacetime trajectory, eventually making some interactions ``unhappen'' while prompting a single one to ``complete its happening,'' until all conservation laws are satisfied over the entire spacetime region.\\

Naturally, the QO phenomenon and the QH hypothesis should be considered separately. Their combination, however, strives to make quantum mechanics more comprehensible and realistic, at the price of admitting that spacetime has some aspects still unaccounted for in current physical theory.\\

\section{``If grandma had wheels, she would be a wagon'': Counterfactual as the essential difference between classical and quantum mechanics} \label{sec:1}

It is the quantum effects themselves, regardless of any interpretation, which are best illustrated by the cynical Yiddish aphorism quoted in this section's title. In classical physics, grandmother's wheels, which she never had, trivially play no role in her dynamics. A quantum grandmother, in contrast, manages somehow to employ them, even to the point of outrunning vehicles. This is the quantum counterfactual \cite{Counterfactuals}, illustrated by the following examples: \\

i) In the Elitzur-Vaidman experiment \cite{IFM}, a bomb is prepared and positioned such that, if struck by an appropriately prepared photon, it will explode. Surprisingly, even when no explosion happens, its mere {\it potentiality} suffices to disturb the photon’s interference: a non-explosive bomb leaves it undisturbed. \\
ii) Hence, in any interaction of a particle with more than one possible absorber, each absorber's capability of absorbing the particle takes part in determining its final position (``collapse''). Consider, e.g., a position measurement of a photon whose wave-function spreads from the source towards a circle of remote detectors: Every detector's {\it non-clicking} incrementally contribute to the photon's eventual position, just like the single detector that does click. \\
iii) Such counterfactuals underlie also quantum non-locality. Bell's theorem \cite{Bell} implies the following counterfactual: Had Bob chosen to measure his particle's spin along an axis other than the one he actually chose, then Alice's spin measurement would give the opposite value.\\

\begin{figure}[h!]
\begin{center}
\includegraphics[scale=0.50]{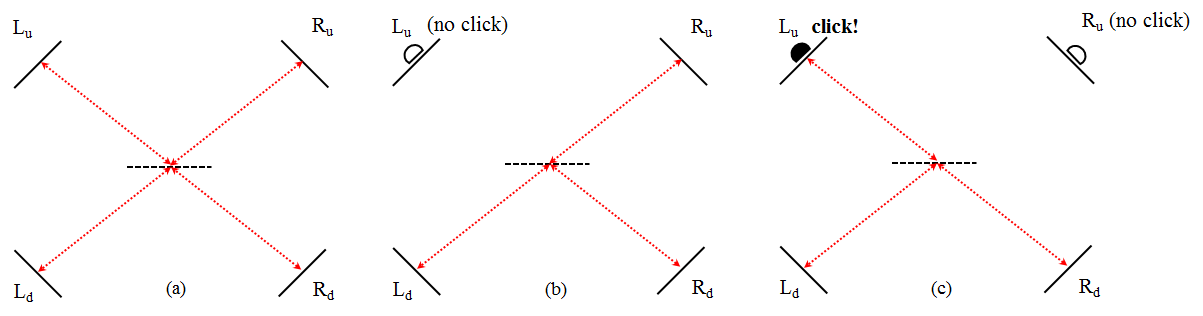}
\caption{ (a) A symmetric interferometer containing a single photon of unknown position and momentum. (b) A non-click on one corner permanently banishes the photon from that corner. (c) That this banishment was an objective effect can be proved by the fact that a second non-click causes its detection in the excluded location.
\label{Fig1}}
\end{center}
\end{figure}

This peculiar status of counterfactuals in QM stems from the uncertainty principle. Consider the simple setting in Fig. \ref{Fig1}. A beam-splitter (BS) is positioned between four equidistant mirrors, Left/Right/Up/Down. We know only that, within the device, a single photon crosses the BS back and forth. This situation, with four possible positions and momenta of the photon, encapsulates the uncertainty $\Delta x\Delta p \ge \hbar/2$, well-known from the standard  Mach-Zehnder interferometer (MZI).

To see that, loosen mirror $L_u$ to enable it to measure the photon's possible interaction with it. In $50\%$ of the cases, the detector will remain silent. This means, with certainty, that the photon hits only $L_d$ and then reflected back to BS, where it splits towards {\it both} $R_u$ and $R_d$, then reflected back, reunited by BS and returns {\it only} to $L_d$. We can therefore leave the detector in $L_u$ and be sure that, despite the photon's endless oscillations, it will {\it never} hit this mirror.
What has happened here? Apparently nothing. A possible event did not occur. Yet the fact that it {\it could} occur endows it a physical say. For if we turn also mirror $R_u$ into a detector, then again in $50\%$ cases it will not click, but now, sooner or later, $L_u$ is bound to click! The second non-event gave rise to a real event, which until that moment, was impossible.

These four possible positions and momenta are typical quantum counterfactuals. As such, they present an acute duality: \\
i) As long as they are not verified, they take an essential causal part in the system's dynamics.\\
ii)	Once, however, one of them is verified, all others vanish without a trace.\\

How, then, can non-events play part in the quantum process? The challenge's acuity is reflected in the various radical moves it has elicited from the reveal interpretations of QM, of which two famous schools have desperately resort to the extreme opposites:\\
{\it Abandonment of ontology: Copenhagen}. Since counterfactuals are facts of our knowledge, just like actual facts, let us define QM as dealing only with knowledge, information, etc., rather than with objective reality. \\
{\it Excess ontology: Many worlds}. The counterfactual does occur, but in a different world, split from ours at the instant of measurement.\\

In what follows we offer an explanation to quantum counterfactuals which is purely physical, derived from quantum theory alone: At the sub-quantum level, grandmother Nature utilizes her proverbial wheels by taking advantage of an otherwise-unhappy consequence of her age: {\it Amnesia}.

\section{When an event is forgotten by Nature herself: Quantum Oblivion} \label{sec:2}
We begin with a very simple yet surprising quantum interaction where one particle emerges from it visibly affected, while the other seems to ``remember'' nothing about it.
\subsection{A non-reciprocal interaction}
\begin{figure}[h!]
\begin{center}
\includegraphics[scale=0.70]{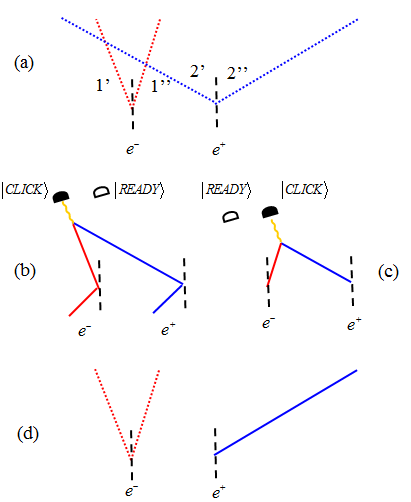}
\caption{An electron-positron interaction and its possible outcomes. (a) The setting. (b-c) Annihilation. (d) Oblivion.
\label{Fig2}}
\end{center}
\end{figure}

Let an electron and a positron, with spin states $|\uparrow_z\rangle=\frac{1}{\sqrt{2}}(|\uparrow_x\rangle +|\downarrow_x\rangle)$ and momenta  $\langle(p_x)_{e^-}\rangle<\langle(p_x)_{e^+}\rangle$, $\langle(p_y)_{e^-}\rangle=\langle(p_y)_{e^+}\rangle$, enter two Stern-Gerlach magnets (SGMs) (drawn for simplicity as beam-splitters) positioned at $(t_0,x_{e^-},y_0)$ and $(t_0,x_{e^+},y_0)$ respectively (see Fig. \ref{Fig2}). The SGMs split the particles' paths according to their spins in the $x$-direction:
\begin{equation}
|\psi_{e^-}\rangle=\frac{1}{\sqrt{2}}(|1'_{e^-}\rangle+|1''_{e^-}\rangle),
\end{equation}
and
\begin{equation}
|\psi_{e^+}\rangle=\frac{1}{\sqrt{2}}(|2'_{e^+}\rangle+|2''_{e^+}\rangle).
\end{equation}
Let technical care be taken such that, if the particles turn out to reside in the intersecting paths, they would meet, at $t_1$ or $t_2$, ending up in annihilation. Two nearby detectors  $|READY_1\rangle$, $|READY_2\rangle$ are set to measure the photons emitted upon pair annihilation, thereupon they would change their states to $|CLICK_1\rangle$ or $|CLICK_2\rangle$.

Let us follow the time evolution of these particles. Initially, at $t_0$, the total wave-function is the separable state:
\begin{equation} \label{initial}
|\psi\rangle=\frac{1}{2}(|1'_{e^-}\rangle+|1''_{e^-}\rangle)(|2'_{e^+}\rangle+|2''_{e^+}\rangle)|READY_1\rangle|READY_2\rangle.
\end{equation}
Depending on the two particles' positions at times $t_1$ or $t_2$, they may (not) annihilate and consequently (not) release photons, which would in turn (not) trigger one of the detectors.
At $t_0\le t < t_1$, then, the superposition is still unchanged, as in \ref{initial}. But at $t_1 < t < t_2$ , either photons are emitted, indicating that the system ended up in
$|1''_{e^-}\rangle)|2'_{e^+}\rangle|CLICK_1\rangle|READY_2\rangle$
{\it or not}, and then
\begin{equation}
|\psi\rangle=\frac{1}{\sqrt{3}}[(|1'_{e^-}\rangle+|1''_{e^-}\rangle)|2''_{e^+}\rangle+|1'_{e^-}\rangle|2'_{e^+}\rangle]|READY_1\rangle|READY_2\rangle,     \end{equation}
which is entangled: one component of it is a definite state, while the other is a superposition in itself.

\begin{figure}[h!]
\begin{center}
\includegraphics[scale=0.70]{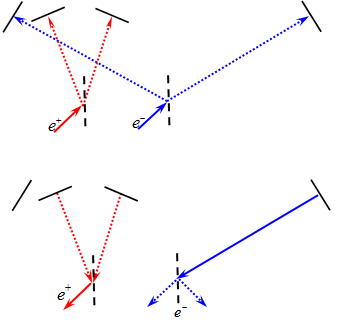}
\caption{The peculiar momentum change due to outcome (d) in Fig. 2: The positron’s interference is disturbed while that of the electron remains intact.
\label{Fig3}}
\end{center}
\end{figure}

Similarly at $t>t_2$: If a photon pair is emitted, we know that the particles ended up in paths $1'$ and $2'$, i.e. 
$|1'_{e^-}\rangle|2'_{e^+}\rangle|READY_1\rangle|CLICK_2\rangle$, otherwise we find the non-entangled state
\begin{equation}
|\psi\rangle=\frac{1}{\sqrt{2}}(|1'_{e^-}\rangle+|1''_{e^-}\rangle)|2''_{e^+}\rangle|READY_1\rangle|READY_2\rangle,     \end{equation}

which is peculiar. The positron is observably affected: If we time-reverse its splitting, it may fail to return to its source.  Its momentum has thus changed. Not so with the electron: It remains superposed, hence its time-reversibility remains intact (Fig. \ref{Fig3}).\\

Summarizing, one party of the interaction ``remembers'' it through momentum change, while the other remains ``oblivious,'' apparently violating momentum conservation. This is Quantum Oblivion (QO).

\subsection{How is momentum conservation restored? The Critical Interval}
It is obviously the intermediate time-interval  $t_1<t<t_2$ that conceals the momentum conservation in QO. The details, however, are no less interesting.

The two particles, during this interval, become entangled in their positions and momenta. Suppose, e.g. that we reunite the two halves of each particle's wave-function through the original BS, to see whether they return to their source. Either one of the particles may fail to do that, on which case the other must remain intact. Similarly for their positions. This is entanglement, identical to that of the electron-positron pair in Hardy's experiment \cite{Hardy}, $|\psi\rangle=\frac{1}{2}(|1'_{e^-}\rangle+|1''_{e^-}\rangle)(|2'_{e^+}\rangle+|2''_{e^+}\rangle)$.

It  is  during  this  interval  that  the  positions  and  momenta  of  the  two  particles  become,
in  contrast  with  ordinary  quantum  measurement,  first entangled  and  then non-entangled (for greater detail of this effect and its implications see \cite{Oblivion}). It remains to be shown that something similar occurs also to the two macroscopic detectors that finalize a experiment. This is done next.
 
\subsection{Oblivion's ubiquity: Every quantum detector's pointer must be superposed in the conjugate variable}
\begin{figure}[h!]
\begin{center}
\includegraphics[scale=0.75]{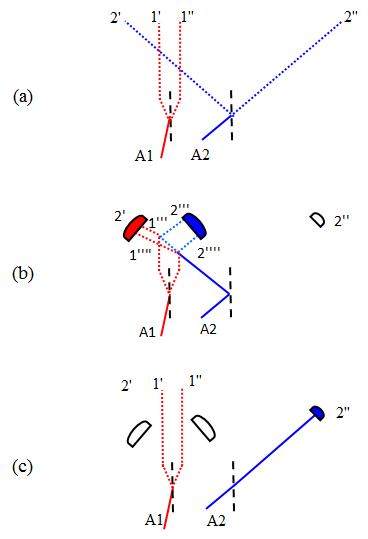}
\caption{Same interaction as in Fig. 2 but with two atoms that do not annihilate upon interaction, hence (a) merely form elastic collision. In this version, the critical interval is the detector's long exposure time which does not allow the precise detection time.  (b) Measurement ending up with collision, where the nearby detectors' widths signify that A1's and A2's positions remain unmeasured. (c) IFM of $A2$'s position.
\label{Fig4}}
\end{center}
\end{figure}

Rather than a curious effect of a specific interaction, Oblivion is present in every routine quantum measurement. Its elucidation can therefore shed new light on the nature of measurement, further enabling some novel varieties thereof. \\
Ordinary quantum measurement requires a basic preparation often considered trivial. Consider e.g., a particle undergoing simple detection (as routinely employed during spin measurements \cite{Wan}). The detector's pointer, positioned at a specific location, reveals the particle's arrival by receiving momentum from it. This, by definition, requires the pointer to have considerable momentum certainty (usually $0$). In return, however, the pointer must have {\it position uncertainty}.
Let this tradeoff be illustrated with a slight modification of our first experiment. In the original version (Fig. \ref{Fig2}), the two possible interactions were annihilations, which were mutually exclusive. For the present purpose, however, let us replace annihilation by mere (elastic) collision (Fig. \ref{Fig4}). In other words, two superposed atoms $A1$ and $A2$ interact like the electron and positron in Fig. \ref{Fig2}, but instead of annihilating, they just collide. This can now happen on {\it both} possible occasions at $t_1$ and $t_2$, namely at the two locations where $A1$ can reside.
With annihilations thus dropped, the outcome is even more interesting. Suppose that the detector on path $2''$ remains silent (Fig. \ref{Fig4}b): We are now certain that the two atoms have collided, but remain oblivious about this collision's location. What we thus measure is an {\it ordinary momentum exchange}: Both atoms' momenta have been reversed along the horizontal axis. Here, oblivion is small, affecting only the two atoms' positions at the time of the collision. Yet, because $A2$ has vanished from the distant location on $2''$, the final outcome is a coarse position measurement of $A2$.

The situation is much more surprising when we {\it fail} to detect $A1$ and $A2$ in the paths to which they would be diverted in case of collision, namely $1'''$, $1''''$, $2'''$, $2''''$. We are now certain that $A2$'s superposition is reduced to path $2''$ while $A1$ has returned to its initial superposition over $1'$ and $1''$ (Fig. \ref{Fig4}c). In other words, $A2$ undergoes momentum oblivion, which is again an ordinary position measurement of $A1$, but this time the measurement is interaction-free, offering a special opportunity to observe how a counterfactual takes an integral part in the quantum evolution.

To summarize: We have studied an asymmetric interaction between two atoms, where two halves of $A1$'s wave-function interact with one half of $A2$. Two momentum exchanges between the atoms can occur: Either \\
i)	$A2$ turns out to have collided with $A1$. This amounts to $A2$ undergoing position measurement. The price exacted by the uncertainty principle is a minor position oblivion of $A1$ and $A2$. \\
Or\\
ii)	$A2$ turns out to have not collided with $A1$. This again amounts to $A2$ undergoing position measurement- collapsing it to the remote $2''$ path. Hence its momentum (interference) is visibly disturbed. This time, however, if $A1$ serves as a detector's pointer, its oblivion is amplified to a macroscopic scale, making the measurement interaction-free. \\

As both (i) and (ii) occur under unusually high space- and time-resolution, they enable a novel study of the Critical Interval. During this interval, entanglement between the two atoms has ensued, as they have assumed new possible locations

\begin{equation}
|\psi\rangle=\frac{1}{2}[|2'''\rangle(|1'''\rangle+|1''''\rangle)+|2''\rangle(|1'\rangle+|1''\rangle)],
\end{equation}

which have remained undistinguished until the macroscopic detectors indicated that no annihilation has occurred. This has broken the symmetry between the two wave-functions, finalizing the interaction and sealing the oblivion.

The generalization is therefore natural: During every quantum measurement, the detector's pointer interacts with the particle in the same asymmetric manner as atoms $1$ and $2$ above: Part of the particle's wave-function with the whole wave-function of the pointer. To make the analogy complete, recall that in reality the pointer's superposition is continuous rather than discrete. As the pointer thus resides over a wide array of locations, momentum measurement becomes much more precise.

\subsection{From IFM to AB: Several quantum phenomena based on QO}
Not only is Quantum Oblivion essential for every quantum measurement, it is also present beneath several well-known variants thereof. In \cite{Oblivion} we have demonstrated QO's underlying IFM \cite{IFM}, Hardy's paradox \cite{Hardy}, weak measurements \cite{AAV,ACE}, partial measurement \cite{ED1,EC1}, the Aharonov-Bohm \cite{AB} and the quantum Zeno \cite{Zeno} effects.
This passage from discrete to continuous superposition also opens the door for several interesting interventions, studied in  \cite{Oblivion}.

\subsection{Summary: Momentum measurement as an example of Oblivion}

Let us summarize the Oblivion effect with the commonest and most basic example. Every time a detector's pointer is set to measure the momentum of a particle that hits it, the following happens.\\
i)	The detector's pointer, prepared with (almost) precise zero momentum, is consequently superposed in space.\\
ii)	Therefore, not only one particle-pointer interaction takes place; rather, several interactions occur, one after another, in all the pointer's superposed locations, as the particle proceeds along them during the Critical Interval.\\
iii) In all these interactions, the photon and the pointer momentarily ``measure'' each other's state, \\
iv)	And each of these mutual ``measurements'' mixes position and momentum, hence being very inaccurate. \\
v)	Yet none of them is amplified to a full measurement. A complex superposition of correlated states thus builds up during the Critical Interval. \\
vi)	Only then does the pointer's state undergo amplification into a full measurement, as follows: All its possible positions fall prey to Oblivion, while all possible momenta add up to a precise value.\\
vii) All the above equally holds for IFM, where the momentum measured is zero.\\
Our counterfactuals, then, become demystified. During the CI, grandmother's wheels do appear and vanish time and again beneath the quantum noise. In the present case, these wheels are the pointer's possible locations. It is only because we choose to measure the pointer's momentum that we give up the options to extract these positions, that have now turned into counterfactuals. Otherwise, as shown in the previous section, had we chosen to measure them instead of the momentum, other quantum surprises would emerge.

\section{Time-symmetric quantum causality: TSVF and weak measurements}
Like two tunnels dug underneath towards one another until merging into one, the above line of investigation turns out to complement another research program, much older and renowned for its surprising predictions, and also verified time and again in laboratories. In what follows we introduce the Two-State Vector Formalism and its offshoot, weak measurement, before proceeding to describe how the two tunnels meet.

\subsection{Why two-state-vectors?}
TSVF originates from the work of Aharonov, Bergman and Lebowitz \cite{ABL}. It asserts that every quantum system is determined by two wave-functions: One (also known as the pre-selected wave-function) evolves forward in time while the other (post-selected) evolves backward. The forward- and backward-evolving wave-functions, $|\psi_i\rangle$ and $\langle \psi_f|$ respectively, define the so called two-state vector $\langle \psi_f|~|\psi_i\rangle$.
The two wave-functions are equally important for describing the present of the quantum system via the weak value of any operator $A$ defined by:
\begin{equation}
\langle A \rangle_w = \frac{\langle \psi_f|A|\psi_i\rangle}{\langle \psi_f|\psi_i\rangle}
\end{equation}
Here a logical catch ensues: {\it ``The state between two measurements'' cannot be revealed by measurement.}
Weak measurement \cite{AAV,ACE} was conceived in order to bypass this obstacle as well as to test other TSVF's predictions. This led to numerous intriguing works, both theoretical and experimental.

\subsection{Anomalous momentum exchanges predicted under special pre- and post-selections}
Apart from its time-symmetry, TSVF reveals even subtler symmetries with respect to values usually considered inherently positive. Consider the surprising values for mass and momentum in the following two experiments \\
i) The Three Boxes Paradox \cite{boxes} is a by-now familiar surprise yielded by TSVF, of which the underlying logic can serve as an introduction to the idea of QH. A particle is prepared with equal probability to reside in one out of three boxes:
\begin{equation}
|\psi_i\rangle=\frac{1}{\sqrt{3}}(|1\rangle+|2\rangle+|3\rangle).
\end{equation}
Later it is post-selected in the state
\begin{equation}
|\psi_f\rangle=\frac{1}{\sqrt{3}}(|1\rangle+|2\rangle-|3\rangle).
\end{equation}
What is the particle's state between these two strong measurements? By definition, {\it projective measurement is unsuitable for answering this question}, as it reveals the state upon the intermediate measurement. It is {\it weak} measurement, again, that comes to help. TSVF predicts the following weak values of the projection operators, $P_i \equiv |i\rangle\langle i|$ for $i=1,2,3$:
\begin{equation}
\begin{array}{lcl}
\langle P_1 \rangle_w = 1\\
\langle P_2 \rangle_w = 1\\
\langle P_3 \rangle_w = -1.
\end{array}
\end{equation}
Therefore, the total number of particles is $1$, as it should be, but it is a sum of two ordinary particles plus one odd. 


The last equation denotes negative weak value for the very exitance in the third box. In order to fully grasp the paradoxical nature of this term, let us consider within the context of its standard versions:\\

If ``probability $1$'' means ``The particle certainly resides within this box''; \\

and ``probability $0$'' means ``The particle has never resided within this box'';\\

then ``probability $-1$'' means ``The particle certainly {\it un}resides within this box.''\\

As absurd as third expression may sound, this is its simplest non-mathematical meaning, the alternative being dismissing it as meaningless. The choice in \cite{boxes} to trust the mathematics, has led to assigning a negative sign to every interaction involving the third box, as long as it is weak enough \cite{potential}. Obviously, this cannot be related to the particle's charge, as it played no role from the beginning. The remaining choice is mass. The simplest way to prove this prediction is through the particle's momentum: A collision with another particle must give the latter a ``pull'' rather than a ``push,'' even though their initial velocities were opposite.

In passing, it is worth comparing this step of the TSVF to Dirac's choice to trust the mathematics upon encountering the negative value for the electron's charge, following the dual solution for his famous equation. That choice has later led to the discovery of the positron. This may be the case with Aharonov's present choice as well.

\begin{figure}[h!]
\begin{center}
\includegraphics[scale=0.60]{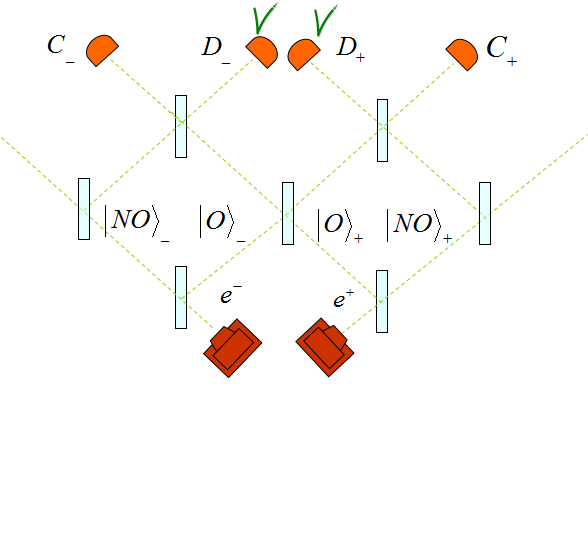}
\caption{Hardy's setup.
\label{Fig5}}
\end{center}
\end{figure}

Can this prediction be put to test? A preliminary version has been carried out by Steinberg's group \cite{Stein} \\

ii) The importance of negative weak values becomes highly visible through Hardy's experiment \cite{Hardy,WHardy,AC}. Two MZIs overlap in one corner (See Fig. \ref{Fig5}). They are tuned such that electron entering the first MZI will always arrive at detector $C_-$, while a positron entering the second MZI will always arrive at detector $C_+$. Therefore, when the electron and positron simultaneously traverse the setup, they might annihilate at the intersection or make their partner reach the ``forbidden'' detector $D_- / D_+$ for the electron/positrion respectively. In case no annihilation was detected we can exclude the case they booth took the overlapping path $O$ and their state becomes
\begin{equation} |\psi_i\rangle=\frac{1}{\sqrt{3}}(|O\rangle_+|NO\rangle_-+|NO\rangle_+|O\rangle_-+|NO\rangle_+|NO\rangle_-),
\end{equation}
{\it i.e.}, at least one of the particles took the non-overlapping ($NO$) state.
The interferometers were tuned such that $C_-$ clicks for the electron constructive interference state $\frac{1}{\sqrt{2}}(|O\rangle_-+|NO\rangle_-)$, $D_-$ clicks for the position constructive interference state $\frac{1}{\sqrt{2}}(|O\rangle_- - |NO\rangle_-)$ state, and similarly for $C_+$ and $D_+$.
Therefore, observing clicks at $D_-$ and $D_+$ is equivalent to post-selection of the state
\begin{equation}
|\psi_f\rangle=\frac{1}{2}(|O\rangle_+ - |NO\rangle_+) (|O\rangle_- - |NO\rangle_-).
\end{equation}
This post-selection is rather peculiar since a click at $D_-$ naively tells us that the positron took its overlapping path, while click at $D_+$ naively tells us that the electron took its overlapping path. This scenario, however, is impossible, because we know annihilation did not occur. 

The paradox can be resolved within the TSVF. When we calculate the weak values of the various projection operators we find out that
\begin{equation}
\langle\Pi_{O}^- \Pi_{O}^+\rangle_w=0
\end{equation}
and
\begin{equation}
\langle\Pi_{NO}^- \Pi_{O}^+\rangle_w=\langle\Pi_{O}^- \Pi_{NO}^+\rangle_w=+1,
\end{equation}
while
\begin{equation}
\langle\Pi_{NO}^- \Pi_{NO}^+\rangle_w=-1.
\end{equation}
This leads us to conclude that although the number of pairs is $1$, we have two ``positive'' pairs and one ``negative'' pair- a pair of particles with opposite properties. The pair in the ``NO-NO'' path creates a negative ``weak potential'' \cite{Potential}, that is, when weakly interacting with any other particle in the intermediate time, its effect will have a negative sign.
Moreover, we can see now the cancelation of positive and negative weak values by considering projections on the non-overlapping paths:
\begin{equation}
\langle\Pi_{NO}^- \rangle_w= \langle\Pi_{NO}^- \Pi_{O}^+\rangle_w+\langle\Pi_{NO}^- \Pi_{NO}^+\rangle_w=1-1=0
\end{equation}
\begin{equation}
\langle\Pi_{NO}^+ \rangle_w= \langle\Pi_{O}^- \Pi_{NO}^+\rangle_w+\langle\Pi_{NO}^- \Pi_{NO}^+\rangle_w=1-1=0
\end{equation}

iii) We can now derive from these experiments precisely the conclusion needed for our thesis: {\it The negative values that they reveal exist also in ordinary measurements, only perfectly hidden}.\\
Consider then, a photon split by a BS, into $1/3$ and $2/3$ beams, then split the $2/3$ beam further into two $1/3$ beams (this setup was recently suggested by Vaidman to answer the question ``where have the photons been?'' \cite{past1,past2,past3}. Perform weak measurements on all three thirds of the wave-function. Now delay all three parts in separate boxes. This delay enables you to choose between two options: Either \\
a)	Reunite the three beams for interference, and post-select destructive interference with negative sign of the third. You will get the results of in the above three boxes experiment, namely, two photon and one negative-mass photon.\\
Or \\
b)	Reunite only the two split $1/3$s back into the original $2/3$ beam, then perform strong position measurements on each beam. You will get a photon on one side and nothing on the other. No post-selection can reveal nothing unusual in the earlier weak measurements.
What was so far a trivial ``nothing'' thus turns out to be something much more profound, namely, a perfect mutual cancellation of positive and negative masses. Similarly for the real photon detected on the other half: It turns out to be the sum of two odd weak values.

\section{The synthesis: How does grandma dispose of the wheels she denies she ever had?}
So far, our discussion went strictly within standard quantum formalism. Oblivion straightforwardly follows under a finer time-resolution of the quantum interaction. Encouraged by TSVF, we now take a step further asking: How does oblivion evolve in space and time?
This is admittedly a dissembling question, pretending not to know that quantum measurement itself still poses unresolved issues; that ``collapse of the wave-function'' is hotly debated, etc. We thus enter the daunting minefield of interpretations of QM, where each choice carries its penalty. Equally, however, these difficulties may serve as positive incentives: To the extent that QO and TSVF illuminate the nature of quantum measurement, it is worth rephrasing older questions in the terms of these frameworks.

\subsection{Grandma's wheels revisited: How do counterfactuals vanish without a trace?}
Our proposal focuses on aspect (ii) of the quantum counterfactuals' dual nature (Section 2), namely, their perfect self-annihilation. Recall that any possibility to observe a counterfactual after the measurement, such as retrieving information about a particle's momentum after its position was measured, entails straightforward violation of the uncertainty principle.
The challenge, therefore, is to explain not only how the counterfactual exerts such an essential role, but also how, having done that, it manages to disappear without a trace.
\subsection{Clue 1: Counterfactuals momentarily show up prior to Oblivion}
The first clue is already in our possession: QO has shown that counterfactuals can be observed, although indirectly beneath great quantum noise, during the Critical Interval (Sec. 3.2).
\subsection{Clue 2: Time-symmetric causality may be the source of quantum oddities}
We now return to Aharonov's Two State-Vector formalism described in Sec. 4.1 above, where causality is highly time-symmetric at the quantum level. Elsewhere \cite{ACE,future} we have explained our endorsement of this approach. First, it is appealingly simple. Take for example the EPR experiment: One experimenter's choice of measurement obviously affects the outcome of another measurement chosen at that moment, far away, by the other experimenter. This appears disturbing, however, only as long as one assumes the familiar causation, going from the past emission of the particles to the later measurements. If, in contrast, one allows causal relations to somehow go back and forth between the past event and its successors, much of the mystery dissolves. {\it What appears to be nonlocal in space, becomes perfectly local in spacetime}. Similar explanations, equally elegant, await many other quantum paradoxes, from Schr\"{o}dinger's cat to the Quantum Liar paradox \cite{ED2}.

To be sure, the above ``somehow'', by which causality runs along both time directions, still needs much clarification. The model also takes the price of admitting that there is something to spacetime that still lies beyond physics' comprehension. These issues will be discussed in consecutive works. Yet we find the advantages of time-symmetric quantum causality compelling enough to pursue them, even before they mature into a fully-fledged physical theory.
Our second clue is therefore: {\it Quantum peculiarities can be better explained by allowing causality to be time-symmetric at the quantum level}.
\subsection{Clue 3: Time-symmetric quantum causality entails unusual physical values}
While time-symmetric causality has been conceived by several quantum physicists (most notably Cramer's Transactional Interpretation \cite{transactional}), surprising experimental derivation of it are solely due to the TSVF. These are the odd weak values described in Section 4.2 above. Although they could be derived from standard quantum theory, the fact is that they never crossed anybody's mind unless guided by TSVF. Paradoxically, it is these exotic features that go beyond interpretation, having won several laboratory verifications worldwide.
Here then is our third clue: {\it Unusual quantum values that appear to be mere noise due to the pointer's uncertainty, turn out to be physically real, revealed by weak measurements. They may be unusually large, unusually small, and negative, even in the case of the mass, which in classical physics is known to be only positive}.
\subsection{Synthesis: Quantum Hesitation}
We have what we need. The established Oblivion effect (Section 3), plus the hypothesis of time-symmetric quantum causality (Section 4.1), plus the latter's more specific discovery of odd values (Section 4.2), merge into the following coherent picture:
Oblivion is an evolution not only of states, but of histories. ``Collapse,'' whereby quantum counterfactuals first play an essential role and then succumb to oblivion, occurs by (i) retarded and advanced actions complementing one another along the same spacetime trajectory, and (ii) negative values taking place alongside with positive ones.
To prove this assertion, we have to show that its most exotic ingredient, namely the odd values, rather than being fringe phenomena, are in fact omnipresent. As paradoxical as this may sound, the mathematical proof is fairly straightforward.
\subsection{Ubiquity: Weak values affect all Quantum Values}
As in the case of oblivion, we now show that weak values, rather than an exotic curiosity, are part and parcel of every quantum value. The proof relies on two ranges of continuum. Suppose that we gradually move (i) from weak measurements to strong ones, and/or (ii) from special pre- and post-selections to non-special ones. Would odd weak values vanish? This indeed appears to be the case, but in reality they become stronger, only counterbalanced by opposite weak values so as to give the familiar quantum values. Moreover, as one of us has shown in \cite{discrimination} there exists a continuous mapping between weak and projective measurement, allowing to decompose the latter into a long sequence of weak measurements. Performing a sequence of weak measurements on a single particle, is amount to a biased random walk with incremental. Finally, the measured state is randomly driven into a definite vector out of the measurement basis.
 
Ergo, Upon gradually moving from weak to strong measurements, odd values do not diminish, but rather increase. Its only through their addition with other weak values that they add up or cancel out, either completing or canceling each other to give the familiar quantum values, including $0$.

\subsection{The missing link: Weak values play an essential role in quantum interference and contextuality}
The above idea seem to be independently emerging in the very current literature on weak values. The main argument of this work nothing short of astonishing, but perfectly converges with ours: Weak values take part in the formulation of the most salient features of the quantum realm.\\
i)	Contextuality: When you measure a particle's spin along a certain direction and get the value ``up,'' you can be sure that, had the SGM been oriented to the opposite direction, the results would have been ``down.'' Similarly for other directions: The outcome should deviate from your ``up'' by the angle difference between the directions. This is common logic. Quantum contextuality, however, poses an odd restriction on these counterfactual outcomes: They have no objective existence, because they depend on the very orientation you chose first. In other words, there cannot be some objective ``up/down'' along a certain spin direction which has existed prior to your choice, such that the angle you choose would be related to it. On the contrary, it is the direction that you chose for whatever reason that serves as the basis for all other possible choices. 
Although contextuality and non-locality of quantum phenomena are dominating strong values, the work of Pusey \cite{Pusey}, as well as our previous works \cite{future,AC} have recently shown respectively that interfering weak values can well account for the contextuality and non-locality of QM. \\
ii)	Interference: Classical waves interfere in strict accordance with everyday intuition. Single-particle interference, in contrast, is paradoxical in that the location of a single particle after passing the interference device is strictly determined by counterfactual paths.
Here too, Mori and Tsutumi \cite{WI} have recently argued that weak values take part in the formation of this quantum phenomenon. Their ``weak trajectories,'' which the single particle is supposed not to have traversed, are shown on one hand to be detectable by weak measurement and on the other hand to add up to give the wave-function's familiar undulatory motion.

\section{Summary: Silence and noise as the progenitors of the quantum signal}
This article recounts a search for a better understanding of the unique causal efficacy of quantum events that failed to occur. We have revisited two opposite characteristics unique to quantum physical values. In terms of information theory, classical values constitute, by definition, signals. Quantum values, in contrast, are sometimes equivalent to lack of signals, namely, silence, and sometimes to a surplus of signals, namely, noise. Perhaps, then, instead of ignoring the former and trying to filter out the latter, it is time to take them as complementary to signals, even as their constituents? We have therefore studied two such quantum phenomena.\\
{\it Interaction-Free Measurement} indicates that a detector's silence is as causally potent as its click. Seeking to understand this potency, we pointed out Quantum Oblivion as the basic mechanism underlying it. Upon further study, QO turned out to underlie every standard quantum measurement.\\
{\it Weak Values} appear amid enormous quantum noise inflicting the quantum weak measurement. TSVF, however, has extracted from them a great deal of novel information about the quantum state. They reveal, e.g. negative weak values emerging in many quantum states. Recently, it has been argued that these weak values take part also in more common quantum phenomena, such as contextuality and interference.\\

Our synthesis of these two lines of research has led to the model of Quantum Hesitation: {\it The weak values, taking part in the time-symmetric evolution of the quantum state, are responsible for the very phenomenon of quantum measurement which we take as real}. Following the TSVF, we suggest that the measurement's outcome, randomly coming out of several possible ones, is the sum of the two state vectors going back and forth in time. This outcome may therefore emerge from weak values, even odd ones, brought by these two state vectors. Similarly for the even more intriguing disappearance of the other possible outcomes: It may be negative weak values, contributed by one state vector, which precisely cancel the positive weak value of the other state vector. Hence, perhaps, the apparently innocuous ``$0$,'' which nevertheless endows quantum grandmother such intriguing mobility.\\
Can this account boast greater rigor than just one more interpretation of QM? In conclusion we want to point out what we consider to be the greatest difficulty hindering such an advance. The above account invokes two distinct causal chains contacting between past and future events, {\it traversing the same spacetime trajectory}. Such a far-reaching idea may invoke more new questions than answers for the old ones. The most acute problem concerns the nature of time: Is there more to physical time then the geometrical characteristics assigned to it by special and general relativity? Or does it allow some yet unknown ``becoming'' \cite{Davies} that involves a privileged ``now'' moving from past to future?\\
The two major time-symmetric interpretations of quantum mechanics, namely TSVF and the Transactional Interpretation, openly admit that these issues are sorely obscure. Recently both schools produced novel accounts that deal with time's nature in radical ways \cite{Kastner,each}. As our own contribution to these issues has been only to stress their acuity \cite{EC2}, we can only hope for this kind of theorizing to be even more radical in the future.

\hfill

\ack{We thank Tomer Landsberger and Tomer Shushi for very helpful comments and discussions. E.C. was supported in part by Israel Science Foundation Grant No. 1311/14.}

\medskip

\hfill
\hfill


\begin{thebibliography}{1}

\bibitem{Counterfactuals} Vaidman L 2009 Counterfactuals in quantum mechanics, in
Compendium of Quantum Physics: Concepts, Experiments, History and Philosophy Greenberger D, Hentschel K and Weinert F (eds.) (Berlin: Springer-Verlag) p 132 ({\it Preprint} arXiv:0709.0340)

\bibitem{IFM} Elitzur A C and Vaidman L 1993 {\it Found. Phys.} {\bf 23} 987

\bibitem{Bell} Bell J S 1964 {\it Physics} {\bf 1} 195

\bibitem{Hardy} Hardy L 1992 {\it Phys. Rev. Lett.}  {\bf 68} 298

\bibitem{Oblivion} Elitzur A C and Cohen E 2015 {\it Int. J. Quantum Inf.} {\bf 12} 1560024

\bibitem{Wan} Wan K K and McLean R  G 1991 {\it J. Phys. A: Math. Gen.} {\bf 24} L425

\bibitem{AAV} Aharonov Y, Albert D and Vaidman L 1988 {\it Phys. Rev. Lett.} {\bf 60} 1351

\bibitem{ACE} Aharonov Y, Cohen E and Elitzur A C 2014 {\it Phys. Rev. A} {\bf 89} 052105

\bibitem{ED1} Elitzur A C and Dolev S 2001 {\it Phys. Rev. A} {\bf 63} 062109

\bibitem{EC1} Elitzur A C and Cohen E 2011 {\it AIP Conf. Proc.} {\bf 1408} 120

\bibitem{ED2} Elitzur A C and Dolev S 2006 {\it AIP Conf. Proc.} {\bf 863} 27


\bibitem{AB} Aharonov Y and Bohm D 1959 {\it Phys. Rev.} {\bf 115} 485

\bibitem{Zeno} Sudarshan E C G and Misra B 1977 {\it J. Math. Phys.} {\bf 18} 756

\bibitem{ABL} Aharonov Y, Bergman P G and Lebowitz J L 1964 {\it Phys. Rev.} {\bf 134} 1410

\bibitem{boxes} Aharonov Y and Vaidman L 1991 {\it J. Phys. A: Math. Gen.} {\bf 24}, 2315  

\bibitem{potential}
Aharonov Y, Cohen E and Ben-Moshe S 2014 {\it EPJ Web Conf.} {\bf 70} 00053

\bibitem{Stein} Resch  K J, Lundeen J S and Steinberg A M 2004 {\it Phys. Lett. A} {\bf 324} 125

\bibitem{WHardy} Aharonov  Y, Botero A, Popescu S, Reznik B and Tollaksen J 2002 {\it Phys. Lett. A} {\bf 301} 130

\bibitem{AC}
Aharonov Y and Cohen E 2015 Weak values and quantum nonlocality, to be published in ``Quantum Nonlocality and Reality'', Mary Bell and Gao Shan (eds.), CUP, ({\it preprint} http://www.ijqf.org/wps/wp-content/uploads/2015/01/AharonovCohen.pdf)

\bibitem{past1} Vaidman L 2013 {\it Phys. Rev. A} {\bf 87} 052104

\bibitem{past2} Danan A, Farfurnik D, Bar-Ad S and Vaidman L 2013 
{\it Phys. Rev. Lett.} {\bf 111} 240402  

\bibitem{past3}  Vaidman L 2014 {\it Phys. Rev. A} {\bf 89} 024102

\bibitem{future} Aharonov Y, Cohen E and Elitzur A C (2015) {\it Ann. Phys.} {\bf 355} 258

\bibitem{transactional} Cramer J G 1986 {\it Rev. Mod. Phys.} {\bf 58} 647

\bibitem{Pusey} Pusey M F 2014 {\it Phys. Rev. Lett.} {\bf 113} 200401

\bibitem{WI}
Mori T and Tsutsui I 2014 Weak value and the wave-particle duality ({\it Preprint} arXiv:1410.0787)


\bibitem{discrimination} Tamir B, Cohen E and Priel A 2015 {\it Quantum Stud.: Math. Found.} {\bf 3}


\bibitem{Davies} Davies P 1996 {\it About time: Einstein's unfinished revolution} (New-York: Simon and Schuster)

\bibitem{Kastner} Kastner R E 2012 {\it The transactional interpretation of quantum mechanics: the reality of possibility} (Cambridge University Press)
    
\bibitem{each} Aharonov Y, Popescu S and Tollaksen J 2014 Each instant of time new universe in {\it Quantum Theory: A Two-Time Success Story} Struppa D C and Tollaksen J M (eds.) (Milan: Springer) pp 21-36
    
\bibitem{EC2} Cohen E and Elitzur A C 2014 {\it EPJ Web Conf.} {\bf 71} 00028

\end{thebibliography}
\end{document}